\documentclass[twocolumn]{aa}
\usepackage{graphicx}
\usepackage{txfonts}
\begin{document}
\title{The evolution of early--type galaxies at $z$$\sim$1 from the K20 survey\thanks{Based on
   observations collected at the European Southern Observatory, Chile (ESO Programme
   70.A-0548), and on observations with the NASA/ESA {\it Hubble
   Space Telescope}, obtained at the Space Telescope Science Institute,
   which is operated by AURA, Inc., under NASA contract NAS 5--26555.}}


   \author{S. di Serego Alighieri\inst{1}
          \and J. Vernet\inst{1}\thanks{\emph{Present address:}ESO,
		  Karl-Schwarzschild-Str. 2, D-85748 Garching bei M\"unchen,
		  Germany}
		  \and A. Cimatti\inst{1}
		  \and B. Lanzoni\inst{2}
		  \and P. Cassata\inst{3}
		  \and L. Ciotti\inst{4}
		  \and E. Daddi\inst{5}
		  \and M. Mignoli\inst{2}
		  \and E. Pignatelli\inst{6}
		  \and L. Pozzetti\inst{2}
		  \and A. Renzini\inst{7}
		  \and A. Rettura\inst{7,8}
		  \and G. Zamorani\inst{2}
          }

   \offprints{S. di Serego Alighieri}

   \institute{INAF -- Osservatorio Astrofisico di Arcetri,
			Largo E. Fermi 5, I-50125 Firenze, Italy\\
              \email{sperello@arcetri.astro.it}
         \and INAF -- Osservatorio Astronomico di Bologna, Via Ranzani
		  1, I-40127 Bologna, Italy
         \and Dipartimento di Astronomia, Universit\`a di Padova,
		 Vicolo Osservatorio 2, I-35122 Padova, Italy \and Dipartimento di Astronomia, Universit\`a di Bologna, Via Ranzani 1, I-40127 Bologna, Italy \and NOAO, 950 North Cherry Avenue, P.O.Box 26732, Tucson, AZ 85726, U.S.A.
		 \and INAF -- Osservatorio Astronomico di Padova, Vicolo
		 Osservatorio 2, I-35122 Padova, Italy
		 \and European Southern Observatory, Karl-Schwarzschild-Str. 2,
		 D-85748 Garching bei M\"unchen, Germany
         \and Universit\'e Paris-Sud 11, Rue Georges Clemenceau 15, Orsay, F-91405, France
		 }

   \date{Received 1 April 2005 / Accepted    }

   \abstract{We have performed VLT spectroscopy of an almost complete sample of 18
early--type galaxies with $0.88\leq z\leq 1.3$ plus two at $z=0.67$,
selected from the K20 survey, and derived the velocity dispersion for 15+2 of
them. By combining these data with HST and VLT images, we study the Fundamental
Plane (FP), the Faber--Jackson and the Kormendy relations at $z$$\sim$1, and compare
them with the local ones. The FP at $z$$\sim$1 keeps a remarkably
small scatter, and shows both an offset and a rotation,
which we interpret in terms of evolution of the mass--to--light ratio, and
possibly of the size.
We show evidence that the evolution rate depends on galaxy 
mass, being faster for less massive galaxies. We discuss
the possible factors driving the evolution of spheroids and compare our results with
the predictions of the hierachical models of galaxy formation.

   }

\titlerunning{The evolution of early--type galaxies at $z$$\sim$1.}
\authorrunning{S. di Serego Alighieri et al.}
   \maketitle
%
\def\re{R_{\rm e}}
\def\msol{{\cal M}_{\sun}}
\def\kv{${\rm K}_{\rm V}$}

\section{Introduction}

The K20 survey (Cimatti et al. \cite{cim02c}) with its selection in the near 
infrared and its very high spectroscopic redshift completeness (92\%) provides 
an unprecedented tool for studying the history of galaxy mass assembly in the 
Universe, and in particular the formation of massive galaxies. In the
currently popular $\Lambda$CDM scenario, massive galaxies are the product of
rather recent hierarchical merging of pre--existing disk galaxies taking
place largely at $z<1.5$ and with moderate star formation rates (e.g.
Kauffmann et al. \cite{kau93} and Cole et al. \cite{col00}). Therefore in
such hierachical merging scenario fully assembled massive galaxies with
${\cal M}\geq 10^{11}{\cal M}_{\sun}$ and with evolved stellar populations
should be very rare at $z\geq 1$.
The alternative possibility is that massive systems formed at much higher
redshifts (e.g. $z\geq 3$), through a short and intense period of star
formation, followed by passive evolution of the stellar population. Such a
possibility is supported by the properties of local and intermediate
redshift spheroids (Renzini \cite{ren99}), and by the existence of old and passive
galaxies at $z$$\sim$1$\div$2 (e.g. Cimatti et al. \cite{cim02a}, Cimatti et al.
\cite{cim04}, McCarthy et al. \cite{mcc04}, Saracco et al.
\cite{sar05}).
The controversy can then be solved by finding a number of evolved massive
galaxies at $z\geq 1$. The problem so far has been the reliable
determination of the mass, since the stellar masses estimated using the
mass to infrared luminosity ratio or the fit to the multicolor spectral
energy distribution are model dependent (IMF) and subject to various
degeneracies (age -- metallicity -- extinction).

The advent of 8--10m class telescopes has allowed dynamical masses to be
estimated more reliably by kinematic studies of galaxies at $z$$\sim$1. Various
groups (e.g. van Dokkum \& Stanford \cite{van03}, Treu et al. \cite{tre05b},
Holden et al. \cite{hol05}, Gebhardt et al. \cite{geb03}, van der Wel et
al. \cite{van05} and references therein) have conducted kinematic studies of
early--type galaxies, selected by optical morphology, both in the cluster and 
in the field environment up to $z$$\sim$1.3. They have used the Fundamental
Plane (FP, Djorgovski \& Davis \cite{djo87} and Dressler et
al. \cite{dre87}) to analyse the evolution of the ${\cal M}/L$ ratio,
generally finding that the most massive spheroids must have formed
at rather high redshift ($z$$\sim$3). In order to understand the formation and
evolution of early--type galaxies over a broader mass range it is necessary
to extend these studies with different selection criteria and better
completeness.

We  address this problem by making spectroscopic observations of
a sample of high redshift galaxies selected from the K20 survey, and 
by exploiting the large collecting area of
the ESO Very Large Telescope (VLT), enhanced by the exceptionally good red
sensitivity of FORS2 on the VLT.
The selection of the
K20 survey, being in the $K_s$ band, which approximately corresponds to the
rest frame $J$--band at the redshift of our sample, is much more sensitive
to the stellar mass of the galaxy and much less dependent on the possible
presence
of star forming activity than the usual selection in the observed optical
bands.  Therefore it is suited to study the evolution of massive
galaxies in a less biased way.
Furthermore the spectroscopic classification of the K20 survey assigns a
galaxy to the early--type class based on the presence of old stars, definitely a
much more stable property than morphology and therefore less likely to
exclude possible progenitors of today's spheroidal galaxies. In fact the only
progenitors that we might miss are those where the
clear spectroscopic signature of old stars would be overwhelmed by a
major starburst, involving a considerable fraction of the galaxy mass. 
Our sample should therefore be less affected by the so called ``progenitor bias''
(e.g. van Dokkum \& Franx \cite{van01}).
Finally, the fact that the K20 survey and our study cover two independent fields, 
well separated in the sky, helps in reducing the effects of cosmic variance.

We describe here
the results of our kinematic study on the early--type galaxies 
of the sample selected from the K20 survey, while a 
parallel paper (Vernet et al., in preparation) reports on those for disk
galaxies.
We assume a flat Universe with $\Omega_m=0.3$,
$\Omega_{\Lambda}=0.7$, and $H_0=70 {\rm km} {\rm s}^{-1} {\rm Mpc}^{-1}$, and we use
magnitudes based on the Vega system.

\section{Sample selection and observations}

The galaxies to be observed have been selected from the K20 survey,
which contains 545 objects selected in the $K_s$ band, with $K_s$$<$$20.0$,
in two areas of the sky, the CDFS and 0055$-$269 fields covering 32.2 and 19.8 arcmin$^2$,
respectively. We have selected galaxies with spectroscopic redshift between
0.88 and 1.3 and with early--type spectra, i.e. classified as class 1.0
(early--type galaxy without emission lines) and class 1.5 (early--type galaxy
with emission lines) in the original K20 classification, based on a visual
inspection of the spectra (Cimatti et al. \cite{cim02c}).
The limits of the redshift range, particularly the lower one, were set by the need of 
covering with a single spectroscopic setting
the rest--frame wavelength interval between 3700\AA~ and 4500\AA, which contains
the most useful absorption lines for the measurement of the velocity dispersion
and the [OII]3727 emission line doublet. 

Of the 14 and 10 galaxies selected with our criteria in the 
CDFS and 0055$-$269 fields respectively, we have observed
9 and 9 objects (see Table~\ref{sample}), i.e. 75\%.
The remaining objects have been 
excluded for purely practical reasons, i.e. they were either outside of
the mask prepared for multi--object spectroscopy or their position was
conflicting with that of other objects included in the mask. 
A better completeness has
been achieved on the 0055$-$269 field because this, being smaller, is easier
to cover with a single FORS2 mask.
The result of our good completeness
is that the observed galaxies span across the whole range of
luminosities available for K20 early--type galaxies and give an unbiased view of
the properties of early--type galaxies in the selected redshift and magnitude 
ranges. In order to check the properties of galaxies at lower redshift, we
have included in the observed sample two galaxies at $z = 0.67$, one for each
field. Recently a more quantitative classification of the K20 spectra has
become available (Mignoli et al. \cite{mig05}), with some differences with
respect to the original visual
classification. The only practical effect on our sample is that the galaxy
identified as CDFS\_00547 (which is class 1.5 
in Cimatti et al. \cite{cim02c}), with
the recent reclassification would have been excluded from the sample, since 
it is not classified as an early--type galaxy, because its 4000\AA~
break is too small, although the spectral energy distribution is red
(Mignoli et al. \cite{mig05}).
Of the 20 observed galaxies (see Table~\ref{sample}) 11 qualify as
extremely red objects (ERO), according to the $R-K_s>5.0$ definition (Elston,
Rieke \& Rieke \cite{els88}), a confirmation of the power of this colour selection
in identifying evolved galaxies at intermediate redshift.

The selected redshift range excludes the two peaks in the
redshift distribution around $z$$\sim$0.67 and $z$$\sim$0.74 in the K20
survey (Cimatti et al. \cite{cim02b}). Therefore our sample is most likely made
of field galaxies, except for the two galaxies at $z=0.67$, outside our
main redshift range, which probably belong to the structures present at that
redshift in both fields.
The observed sample of galaxies is listed in Table~\ref{sample}. The reader is
referred to the recent release of the K20 survey \footnote{See
http://www.arcetri.astro.it/$\sim$k20/spe\_release\_dec04/index.html.} for the object
identification and for additional information on the individual galaxies.

The spectroscopic observations to obtain the velocity dispersion have
been performed between 31 October and 2 November 2002
with FORS2 on UT4 of the ESO VLT at Paranal. We have used the grism 600z with
slits 1.0 arcsec wide, which gives a resolution $\lambda /\Delta\lambda =
1400$ and covers the wavelength range 7400--10700\AA. The rest--frame
wavelength range for each individual galaxy, although varying
according to its
redshift and its position in the field, includes always the H and K [CaII]
lines and the G band, except for one of the two lower redshift
galaxies (q0055\_00169), for which the spectrum starts around 4000\AA. We used
two multi--object masks, one for each field. The
exposure time was 7.5 and 6.0 hours, and the average seeing was 0.7 and
0.625 arcsec for the CDFS and 0055$-$269 field respectively. For calibration
we have observed the spectrophotometric standard stars Hiltner 600 and
HD49798.

For the morphological and photometric analysis we have used the HST+ACS 5 epochs
mosaicked stacks $F850LP$ images from the Great Observatories Origin Deep
Survey (GOODS, Giavalisco et al. \cite{gia04}) and images from the ESO
Imaging Survey (EIS, NTT+SUSI2 $U,B,V,R$ and NTT+SOFI $J,K_s$) for the CDFS
field, and images obtained with the NTT+SUSI2 ($U,B,V,I$), the NTT+SOFI
($J,K_s$) and the VLT+FORS1 ($R,z$) for the 0055$-$269 field.

\begin {table*}
  \caption{The sample of observed early--type galaxies from the K20 survey}
	\label{sample}
  \begin{tabular}{c c c c c c c c c c c c}
  \hline\hline
  K20 name & $z$ & $R$ & $R-K_s$ & $n$ & $K_V$ & $R_e$ & $\langle\mu^B\rangle_e$ & $\sigma$ & EW[OII] & $M_B$ & $log{{{\cal M}_{{\rm K}_{\rm V}}}\over{\cal M_{\sun}}}$\\
  & & & & & & kpc & mag/arcsec$^2$ & km/s & \AA & & \\
  \hline
  CDFS\_00060 & 1.188  & 24.89 & 5.56 & & & & & & &\\
  CDFS\_00369 & 0.8930 & 23.30 & 4.14 & 1.85$\pm$0.39 & 7.37$\pm$0.18 & 0.84$\pm$0.05 & 17.78$\pm$0.15 & 119$\pm$21 & 9.0$\pm$1.1 & -20.41& 10.31$\pm$0.15\\
  CDFS\_00467 & 0.8956 & 22.90 & 4.63 & 3.45$\pm$0.39 & 5.72$\pm$0.25 & 1.52$\pm$0.10 & 18.40$\pm$0.15 & 140$\pm$18 & 4.3$\pm$0.8 & -21.09& 10.58$\pm$0.11\\
  CDFS\_00468 & 1.019  & 24.86 & 5.03 & & & & & &&\\
  CDFS\_00532 & 1.2115 & 23.92 & 5.22 & 2.04$\pm$0.33 & 6.82$\pm$0.20 & 1.41$\pm$0.10 & 17.43$\pm$0.16 & 260$\pm$30 & 2.5$\pm$0.5 & -21.89& 11.16$\pm$0.10\\
  CDFS\_00547 & 1.2243 & 23.68 & 5.07 & 1.90$\pm$0.31 & 7.50$\pm$0.14 & 0.74$\pm$0.04 & 16.15$\pm$0.14 & 256$\pm$28 & 2.3$\pm$0.4 & -21.76& 10.93$\pm$0.10\\
  CDFS\_00571 & 0.9551 & 22.33 & 4.79 & 5.00$\pm$0.15 & 3.83$\pm$0.02 & 6.81$\pm$0.48 & 20.65$\pm$0.16 & 182$\pm$21 & 3.7$\pm$0.5 & -22.08& 11.23$\pm$0.10\\
  CDFS\_00590 & 1.2235 & 24.35 & 5.28 & 4.98$\pm$0.25 & 4.03$\pm$0.12 & 4.09$\pm$0.52 & 20.54$\pm$0.29 & 119$\pm$49 & 4.1$\pm$1.1 & -21.09& 10.68$\pm$0.36\\
  CDFS\_00633 & 1.0963 & 22.43 & 5.49 & 4.78$\pm$0.21 & 4.00$\pm$0.12 & 6.65$\pm$0.12 & 19.97$\pm$0.05 & 260$\pm$23 & $<1.4$ & -22.71& 11.55$\pm$0.08\\
  CDFS\_00354 & 0.6672 & 22.01 & 4.07 & 3.58$\pm$0.28 & 5.20$\pm$0.20 & 3.08$\pm$0.12 & 20.39$\pm$0.09 & 99$\pm$19 & 2.6$\pm$0.6$^*$ & -20.61 & 10.52$\pm$0.17\\
  \\
  q0055\_00028 & 1.0524 & 24.19 & 5.09 & 0.8$\pm$0.8 & 7.17$\pm$0.48 & 1.17$\pm$0.65 & 18.08$\pm$1.22 & 227$\pm$30 & $<3.6$ & -20.82 & 10.99$\pm$0.26\\
  q0055\_00068 & 1.1043 & 23.53 & 5.18 & 4.3$\pm$1.0 & 4.55$\pm$0.73 & 3.60$\pm$1.00 & 19.75$\pm$0.61 & 84$\pm$31 & 2.3$\pm$0.7 & -21.60 & 10.37$\pm$0.34\\
  q0055\_00114 & 0.8891 & 23.48 & 4.33 & 2.0$\pm$2.0 & 6.71$\pm$1.78 & 2.43$\pm$0.74 & 19.99$\pm$0.66 & 52$\pm$22 & $<2.1$ & -20.50 & 9.98$\pm$0.38\\
  q0055\_00123 & 0.9270 & 22.96 & 4.84 & 3.8$\pm$1.0 & 5.69$\pm$0.67 & 1.23$\pm$0.40 & 17.62$\pm$0.71 & 186$\pm$19 & 1.6$\pm$0.5 & -21.39 & 10.74$\pm$0.16\\
  q0055\_00247 & 0.931  & 24.43 & 4.54 & & & & & & $<2.8$ & &\\
  q0055\_00295 & 1.1680 & 23.75 & 5.59 & 4.8$\pm$0.3 & 4.35$\pm$0.20 & 2.67$\pm$0.70 & 18.93$\pm$0.57 & 170$\pm$28 & $<1.4$ & -21.77 & 10.85$\pm$0.18\\
  q0055\_00318 & 0.8962 & 23.87 & 5.07 & 4.5$\pm$1.1 & 5.17$\pm$0.70 & 1.26$\pm$0.57 & 18.87$\pm$0.99 & 105$\pm$30 & $<0.8$$^*$ & -20.20 & 10.21$\pm$0.31\\
  q0055\_00331 & 0.8980 & 22.95 & 5.09 & 4.9$\pm$0.3 & 4.15$\pm$0.19 & 3.83$\pm$0.77 & 20.08$\pm$0.44 & 216$\pm$46 & $<0.8$$^*$ & -21.40 & 11.18$\pm$0.19\\
  q0055\_00338 & 0.9324 & 23.12 & 4.96 & 2.8$\pm$0.6 & 5.93$\pm$0.50 & 2.41$\pm$0.43 & 19.23$\pm$0.39 & 113$\pm$19 & $<2.5$ & -21.24 & 10.59$\pm$0.16\\
  q0055\_00169 & 0.6699 & 22.70 & 4.23 & 1.3$\pm$0.8 & 7.15$\pm$0.62 & 1.64$\pm$0.36 & 19.50$\pm$0.47 & 90$\pm$19 & $<0.9$$^*$ & -20.14 & 10.32$\pm$0.20\\
  \hline
  \end{tabular}
  $^*$ for [OIII]5007.
\end{table*}

\section{Data Analysis}

\subsection{Morphology}

In order to obtain the effective
radius $R_e$, we have fitted a S\'ersic profile (${\rm log}I(R) \propto
-R^{1/n}$, S\'ersic \cite{ser68}) using GIM2D, a fitting algorithm for
parametric two-dimensional models of surface brightness distribution
(Simard et al. \cite{sim98}).
GIM2D performs a profile fit by deconvolving the data with the point spread
function. We model PSFs with analytic functions from visually selected stars 
in the surrounding ($30'' \times 30''$) region of
each galaxy. We model a different PSF for each region in order to take 
account of PSF variations with the position in the field.
A 2D radial gaussian function has been fitted simultaneously on tens of 
selected stars around the galaxies  of our sample and
outputs have been stacked together to result into a single PSF image for 
each region. A more detailed description of our approach in
modellig GOODS/ACS galaxy morphologies in the  $1.0
<z< 1.5$ range will be given in Rettura et al. (2005, in
preparation).

The results of the bidimensional fit is the semimajor axis $a_{e}$ of the
projected elliptical isophote containing half of the total light, the axis 
ratio $b/a$ and the S\'ersic index $n$, which we have left
as free parameters.
The effective radius is computed from $R_e = a_e
\sqrt{b/a}$ and is given in Table~\ref{sample}.

Magnitudes have been measured on the ground--based images using the SExtractor
(Bertin \& Arnouts \cite{ber96}) BEST estimator. The average surface
brightness within the effective radius (in mag/arcsec$^2$) 
is obtained from the absolute magnitude
M: $\langle\mu\rangle_e = M + 5logR_e + 38.567$, with $R_e$ in kiloparsec.

In order to obtain the morphological parameters in the rest frame $B$ band we have used the
HST+ACS images taken with the F850LP filter and the VLT+FORS1 images
taken with the Gunn $z$ filter, since these are very close to the $B$ band
at the redshift of our galaxies. In any case a small K--correction has
been applied following the prescriptions of Hogg et al. (\cite{hog02})
and Blanton et al. (\cite{bla03}) and using the excellent information
which we have on the spectral energy distribution. The surface
brightness has been corrected for the cosmological $(1+z)^4$ dimming.
We have also corrected for the small Galactic extinction as obtained from 
Schlegel et al. (\cite{sch98}). 

\begin{figure}
\resizebox{\hsize}{!}{\includegraphics{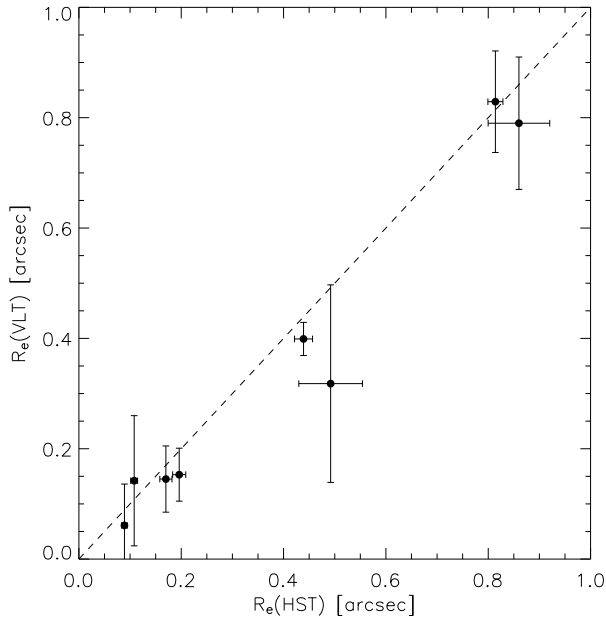}}
\caption{Comparison of effective radii obtained for the CDFS field on the HST+ACS
image and on the VLT+FORS1 one.
}
\label{Figa_e_hst_fors}
\end{figure}

For the 0055$-$269 field we lack HST images, so we have to rely on very
good quality VLT+FORS1 images (seeing 0.62 arcsec). Obviously the
quality of the parameters obtained is worse than for those on the CDFS
field. In order to check that the results are still acceptable and that
the ground--based parameters are not biased, we have repeated the
morphological analysis of the CDFS galaxies on a VLT+FORS1 image
equivalent to that used for the 0055$-$269 field and found that the
parameters obtained are consistent with those based on the HST+ACS
images, although with larger errors (see Fig.~\ref{Figa_e_hst_fors}).

In order to check our consistency with the results obtained by others
fitting the de Vaucouleurs law, we have repeated our fits by fixing the
S\'ersic index to 4, obtaining sizes which are within 40\% of those
obtained with a variable index, some bigger, some smaller, but without a
noticeable dependence on the $n$ value. The difference of the mean sizes
derived with the two different fits is less than 5\%.

\subsection{Spectroscopy}

\begin{figure*}
\centering
\includegraphics[width=17cm]{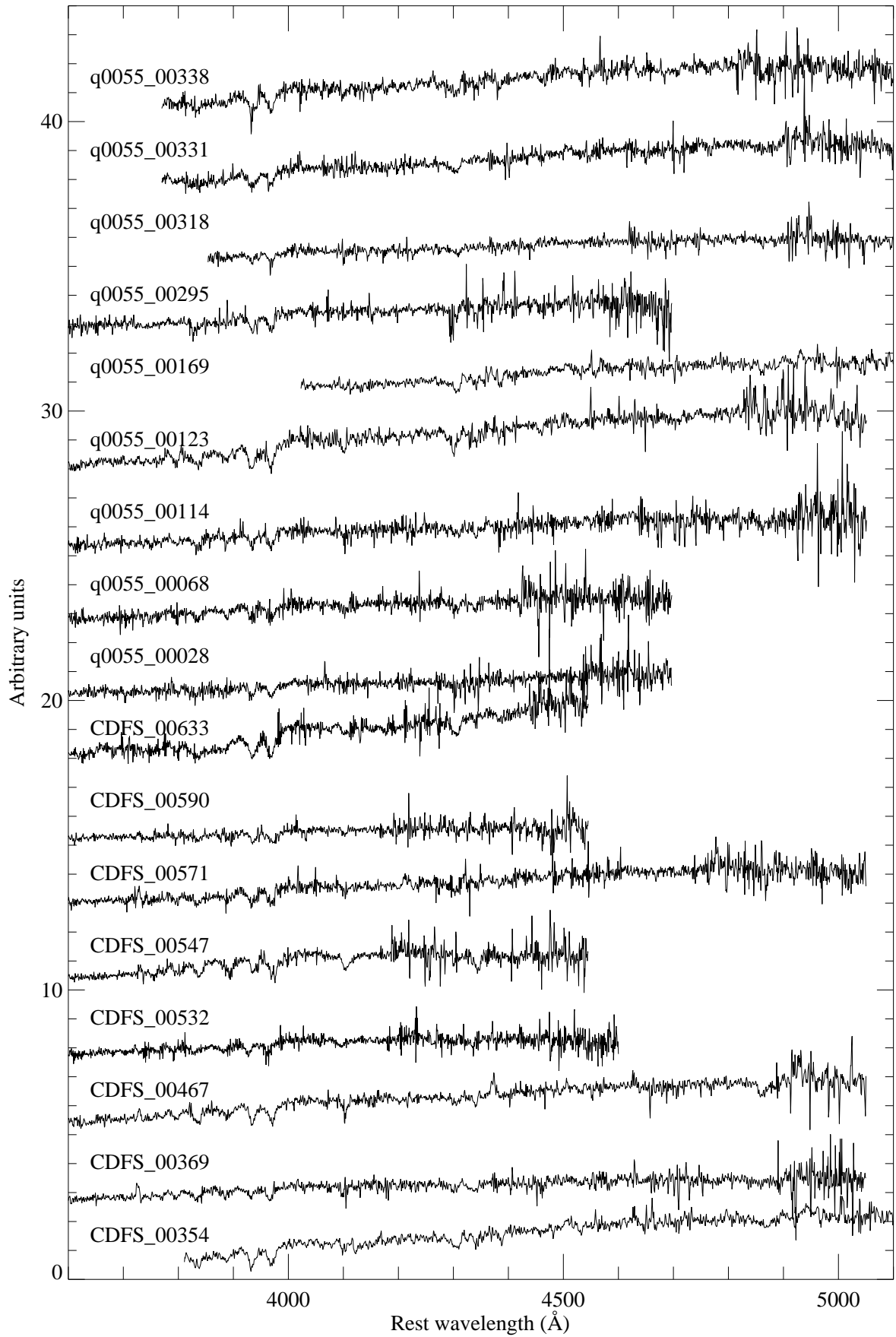}
\caption{The extracted spectra of all the galaxies for which we could obtain
the velocity dispersion.}
\label{Figspectra}
\end{figure*}

The spectra have been reduced using the IRAF software package in the usual way,
including atmospheric extinction correction and flux calibration. For
each galaxy we have extracted a one--dimensional spectrum by coadding a
section centred on the nucleus and one arcsecond long along the slit. This
results in a spectral aperture of 1x1 arcsec$^2$. The extracted spectra are
shown in Fig.~\ref{Figspectra}.

The most noisy -- i.e. sky contaminated -- sections of the spectra have been
masked out and the velocity dispersions have been obtained with the $xcor$
task in IRAF, which uses the cross correlation technique of Tonry and
Davis (\cite{ton79}). The correlation has been performed with template
spectra of stars from the STELIB library (Le Borgne et al.
\cite{leb03}), with spectral types ranging from F6 to K2, by matching their
spectral resolution to that measured on our spectra. We have made
tests with spectra of template stars artificially smoothed to a known
velocity dispersion, and with realistic noise added.
These tests confirm that the results are not biased and
that the quoted errors are realistic. The measured velocity dispersion
has been normalized to a circular aperture with a diameter of $1.19
h^{-1}$ kpc, equivalent to 3.4 arcsec at the distance of the Coma
cluster, following the prescriptions of J\o rgensen et al.
(\cite{jor95b}), as commonly used in FP studies. The changes on the velocity dispersion produced by 
these normalization factors are in any case small, i.e. 6-7\%.
The normalized velocity dispersion is listed in Table~\ref{sample},
except for CDFS\_00060, CDFS\_00468 and q0055\_00247, for which we could
not obtain a reliable value because of the low S/N ratio. 
These three galaxies are the faintest in the $R$ band. The average redshift of the
galaxies with measured velocity dispersion is $z=1.071$ for the CDFS
field and $z=0.983$ for the 0055$-$269 one, excluding the two galaxies at
$z=0.67$.

Since we have noticed that several galaxies show a clear, although weak,
[OII]3727 line emission, we have measured its rest--frame equivalent width 
(see Table~\ref{sample}).
For the four galaxies, for which the [OII] line is outside the observed
wavelength range, we have measured the [OIII]5007 line emission instead, as
noted in the Table.

\section{Results}

\subsection{Morphology}

\begin{figure*}
\centering
\includegraphics[width=17cm]{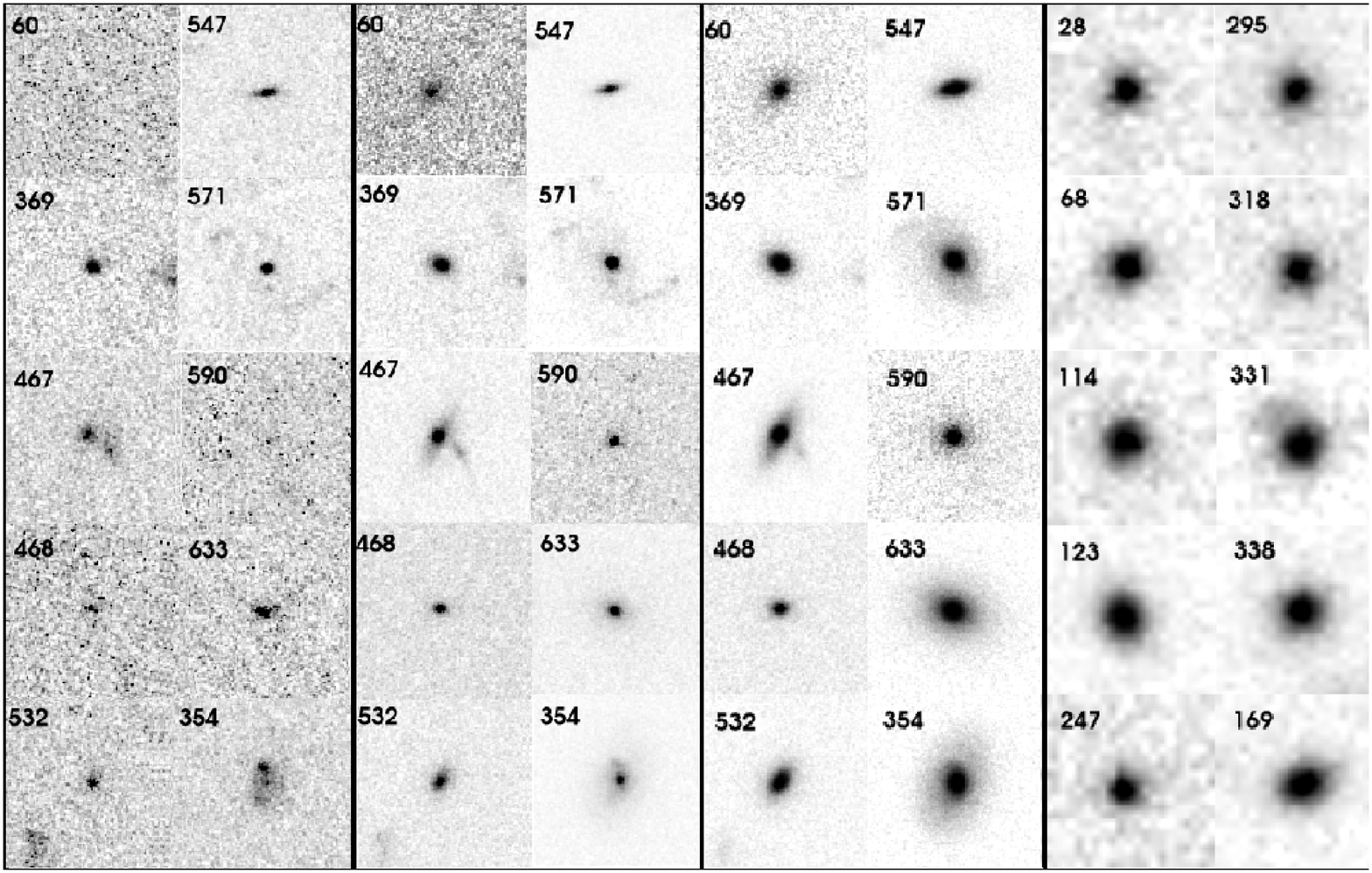}
\caption{Images of our galaxies from the CDFS field (3 panels on the left:
HST+ACS+F435W, HST+ACS+F606W, HST+ACS+F850LP, from left to right) and from the 
0055$-$269 field (rightmost panel,
VLT+FORS1+Gunn z). Each box is $3\times 3$ arcsec. The numbers in the
boxes are the last 3 digits of the object's name (see Table~\ref{sample}).
}
\label{Figmosa_sp1}
\end{figure*}

\begin{figure}
\resizebox{\hsize}{!}{\includegraphics{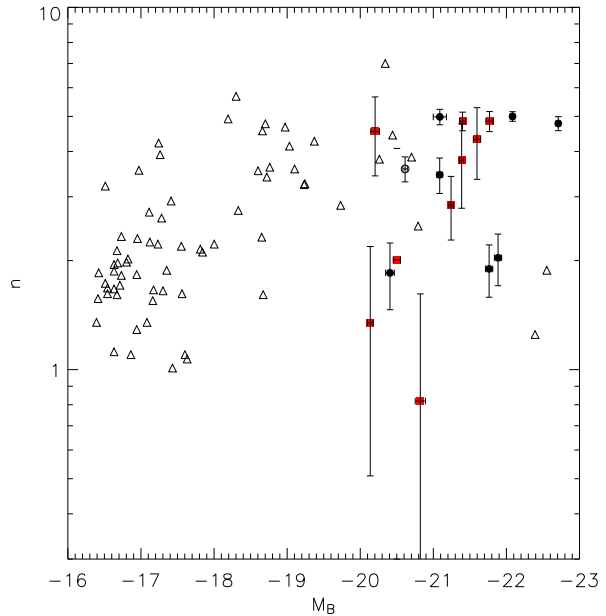}}
\caption{The correlation of the S\'ersic index with the luminosity.
Triangles are the Coma cluster data of Guti\'errez et al.
(\cite{gut04}), black circles and red squares are our $z$$\sim$1 galaxies in the
CDFS and $0055-269$ fields respectively and empty symbols are for
galaxies with $z$$<$$0.88$, while filled ones are for our redshift range
0.88$<$$z$$<$1.3.
}
\label{FignMBvir}
\end{figure}

If one expects that all early--type galaxies should have a
de Vaucouleurs profile, then the S\'ersic indices that we obtain
(see Table~\ref{sample}) appear rather small,
and there are three galaxies in the CDFS and three in the
0055$-$269 field with $n<2.5$.
However also in the local Universe several early--type galaxies are
found to have a small $n$ (see Fig.~\ref{FignMBvir}).
All the galaxies in our sample do have a
prominent spheroidal component, although a disk is visible in a few of them
(see Figure~\ref{Figmosa_sp1}), and all those in the CDFS field have been 
classified as E/S0 in the morphological analysis by Cassata et al. (\cite{cas05}),
except for CDFS\_00571, which is classified as Sa.

Moreover we find that the S\'ersic index correlates with the $B$ band absolute
magnitude (see Figure~\ref{FignMBvir}), as found at low redshift (e.g.
Guti\'errez et al. \cite{gut04}), but with a shift to brighter
luminosities for the same values of $n$.

The distribution of the effective radii for our $z$$\sim$1 galaxies is
shifted towards lower sizes with respect to low redshift samples (see 
Fig.~\ref{FigFP} and Fig.~\ref{FigK}). 
This could be due to a selection effect induced by
the cosmological $(1+z)^4$ dimming in surface brightness, in the sense
that because of the dimming we select higher surface brightness galaxies
which tend to be more compact (see also Sect. 4.5). 

Small sizes might also be due to the presence of an AGN in some of
our galaxies, which could as well explain the large fraction of our objects
showing [OII] emission, particularly in the CDFS field. Therefore we
have checked the presence of our sample objects in the
catalogue of X--ray sources presented by Alexander et al.
(\cite{ale03}), obtained from the deep Chandra observations of the CDFS
field
by Giacconi et al. (\cite{gia02}). We found only one detection:
galaxy CDFS\_00467, which has a 0.5-8 keV X--ray  flux of $2.2\times 10^{-16}$ erg
cm$^{-2}$ s$^{-1}$. For an unobscured type 1 AGN with an X--ray/optical ratio
of 1, the corresponding observed magnitude would be $R$$\sim$26, therefore
contributing about 6\% of the galaxy luminosity. However, since the source
has a hardness ratio of $\approx$0.1, it is probably heavily obscured and much
fainter in the $R$--band. Therefore its influence on the profile should be
completely negligible.

\subsection{The Fundamental Plane}

The FP is a powerful tool to study the evolution of early--type 
galaxies. Because of its dependence on galaxy luminosity,
it is sensitive to recent star formation episodes. 
Therefore the FP could be useful to study the influence
of the environment on the galaxy evolution as, for example, star formation 
induced by merging.

\begin{figure}
\resizebox{\hsize}{!}{\includegraphics{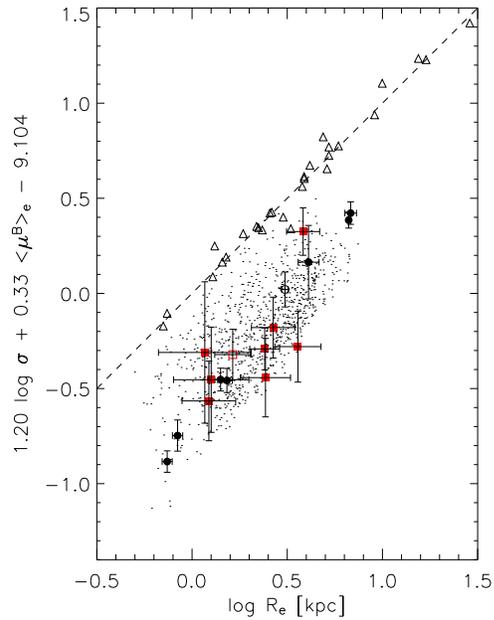}}
\caption{The fundamental plane as obtained by J\o rgensen et al.
(\cite{jor95a},\cite{jor95b}) for the Coma galaxies (triangles). Symbols are
as in Fig.~\ref{FignMBvir}.
The dashed line represents the best--fit plane to the Coma galaxies
as seen edge--on. The small dots represent the GalICS model galaxies (see
Section 5.2).
}
\label{FigFP}
\end{figure}

We show in Figure~\ref{FigFP} the position of our $z$$\sim$1 galaxies on the
FP obtained by J\o rgensen et al. (\cite{jor96}, hereafter JFK96) for the
Coma galaxies at $z=0.0248$ in the $B$--band.
The FP at $z$$\sim$1, compared to the local one, keeps a remarkably small
scatter, particularly for the more accurate CDFS data, and shows a
clear offset, as already noticed previously (e.g. Franx et al.
\cite{fra00}), and likely also a different tilt, suggesting that the
evolution of early--type galaxies depends on their size and/or mass and/or
stellar population. 

In order to parametrize the offset and the rotation we have fitted a plane
to the distribution of our data set, using the expression
(e.g. JFK96):
\begin{equation}
{\rm log} R_e = \alpha {\rm log} \sigma + \beta {\rm log} \langle I\rangle_e + \gamma, 
\label{formFP}
\end{equation}
where $\langle I\rangle_e=L/2\pi R_e^2$, given the definition of the effective radius, is
the average surface luminosity within $R_e$. We
remind the reader that ${\rm log}\langle I\rangle_e = -0.4(\langle\mu\rangle_e-C)$,
where $\langle I\rangle_e$ is in
solar luminosities per square parsec, $\langle\mu\rangle_e$ is in magnitudes per square
arcsecond and the constant $C$ depends on the band, but not on the
cosmology ($C_B$ = 26.982, $C_K$ = 24.982). 

\begin{figure}
\resizebox{\hsize}{!}{\includegraphics{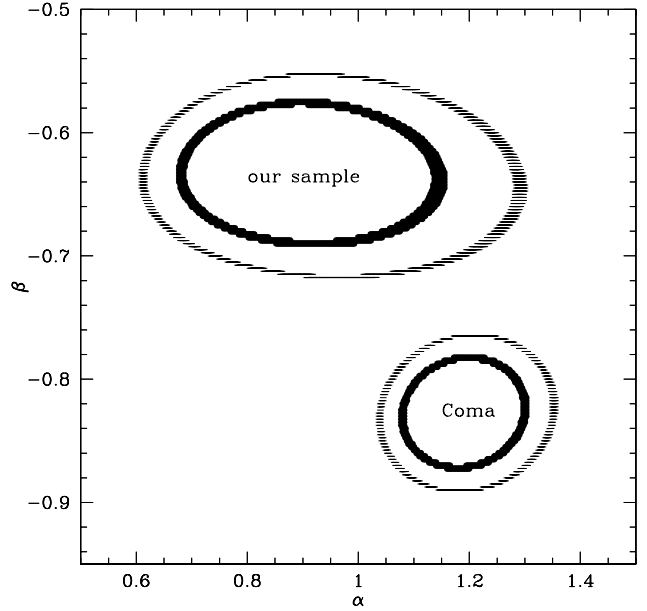}}
\caption{1-sigma (solid) and 90\% (dashed) confidence regions of the
joint distribution of the FP parameters $\alpha$ and $\beta$, for our data--set 
of $z$$\sim$1 galaxies, and for the Coma cluster sample of JFK96.
}
\label{Figokconf}
\end{figure}

The rms scatter of the log$R_e$ residuals from the best--fit FP is 0.11,
while it becomes 0.17 when the Coma cluster parameters are
assumed in the fit, as done in Fig.~\ref{FigFP}.
The best-fitting values of $\alpha$, $\beta$ and $\gamma$, have been derived
by minimizing the sum of the square distances from the plane, weighted by the
observational errors. They are listed in Table~\ref{bestfit}, together with those
obtained for the Coma galaxies of JFK96 by means of the same
procedure.
The 90\% confidence regions of the joint distribution of
$\alpha$ and $\beta$, for our $z\sim 1$ galaxies, and for the Coma cluster
sample of JFK96 (Figure~\ref{Figokconf}) are well separated, showing that no
pair of
($\alpha$, $\beta$) exists that simultaneously provides an acceptable fit to
the two data-sets. As expected, such a result is almost completely due to ACS
data, which have significantly smaller errors for both $R_e$ and $\langle
I\rangle_e$. However, we
have verified that it still holds,
if we use the values of $R_e$ and $\langle I\rangle_e$ obtained from
the de Vaucouleurs ($n=4$) fit to the surface brightness profile.  Thus
our data show that at the 90\% confidence level the FP rotates with
redshift, even if larger samples are needed to better constrain its
evolution.

\begin {table}
\caption{The best fit parameters for the FP in the $B$--band}
\label{bestfit}
\begin{tabular}{c c c c c}
\hline\hline
Sample & Ref. & $\alpha$ & $\beta$ & $\gamma$\\ 
\hline
Coma Cl. & 0 & 1.20$\pm$0.08 & -0.83$\pm$0.03 & -0.17$\pm$0.20\\ 
$0.88<z<1.3$ & 1 & 0.88$\pm$0.16 & -0.63$\pm$0.04 & 0.46$\pm$0.38\\ 
\hline
\end{tabular}

References. 0: JFK96, 1: this work.
\end{table}

In order to study the evolution of the ${\cal M}/L_B$ ratio in a way
consistent with previous work,
we assume that the dynamical mass of the galaxy is (Michard \cite{mic80}):

\begin{equation}
{\cal M} = 5\frac{\sigma^2 R_e}{G},
\label{mass}
\end{equation}

This assumption is equivalent to assuming $R^{1/4}$ homology among
early--type galaxies. We relax this assumption in Section 4.4.

\begin{figure}
\resizebox{\hsize}{!}{\includegraphics{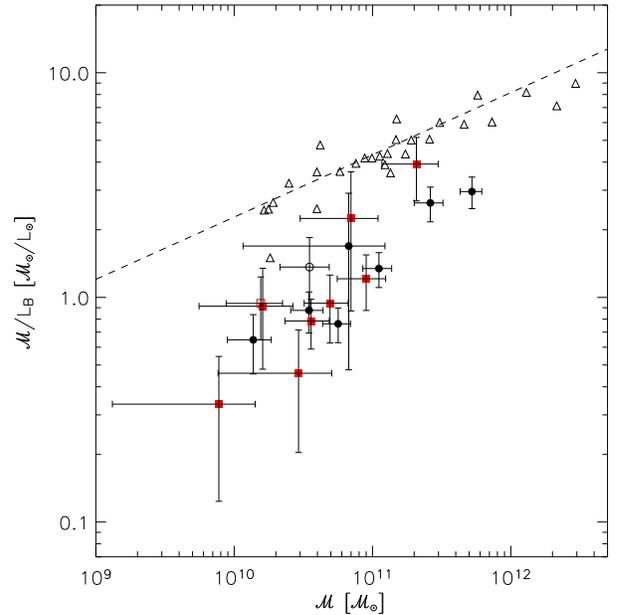}}
\caption{Mass to light ratio vs. mass for our galaxies at $z$$\sim$1 and for
galaxies in Coma. Symbols are as in Fig.~\ref{FignMBvir}.
}
\label{FigML}
\end{figure}

\begin{figure}
\resizebox{\hsize}{!}{\includegraphics{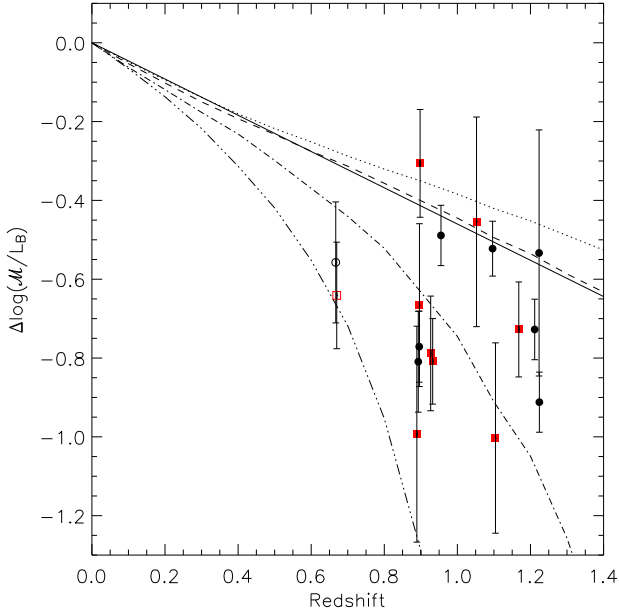}}
\caption{Evolution of the mass to light ratio for our galaxies. Symbols are
as in Fig.~\ref{FignMBvir} and the continuous line is the linear fit
obtained by van Dokkum \& Stanford (\cite{van03}) for 
the evolution of the mass to light ratio of cluster
massive spheroids up to $z=1.27$. The other lines are obtained
with simple stellar population  models using the Chabrier (\cite{cha03}) IMF with a
formation redshift of 5.0, 3.0, 1.5 and 1.0 (top to bottom).
}
\label{FigDeltaML}
\end{figure}

The evolution of the ${\cal M}/L_B$ ratio obtained in this way
is shown in Fig.~\ref{FigML} and Fig.~\ref{FigDeltaML}. The
spheroidal galaxies at $z$$\sim$1 have a brighter $B$--band luminosity for
the same mass than low redshift galaxies by up to a factor of 8, 
particularly for the smaller mass galaxies. 
We note however that our sample goes to smaller masses
than the Coma sample of J\o rgensen et al. (\cite{jor95a},\cite{jor95b}). The
evolution in the ${\cal M}/L_B$ ratio is similar to that of the massive galaxies in clusters for
$\sim 25$\% of our galaxies, while it is stronger for the rest of our sample
(see Fig.~\ref{FigDeltaML}).

Actually, as Treu et al. (\cite{tre05a}) and van der Wel et al. (\cite{van05}),
we find that the evolution is different for galaxies with different mass.
If we fit  a linear slope $\eta$ to the evolution of the ${\cal
M}/L_B$ ratio ($\Delta {\rm log}({\cal M}/L_B) = \eta z$), then we obtain
$\eta = -0.59\pm0.03$ for our 8 galaxies with ${\cal M}>5\times
10^{10}{\cal M}_{\sun}$ and $\eta = -0.86\pm0.05$ for our 9 galaxies
with ${\cal M}<5\times 10^{10}{\cal M}_{\sun}$. These two slopes are
significantly different, showing that the evolution depends on the
galaxy mass (see also Section 5.1). The slope for the more massive
galaxies is close to the value $\eta = -0.46\pm0.04$, 
obtained by van Dokkum \&
Stanford (\cite{van03}) for cluster massive galaxies (${\cal
M}>10^{11}{\cal M}_{\sun}$). In fact we
obtain $\eta = -0.52\pm0.04$ for our 4 most massive galaxies (${\cal
M}>10^{11}{\cal M}_{\sun}$).

Since our sample is selected in the $K_s$ band (approximately rest--frame
$J$--band), the evolution of the ${\cal M}/L_B$ ratio can hardly be due to
selection effects, as could be in the case of selection in the
observed optical bands, which for $z$$\sim$1 are very sensitive to star
formation.

\subsection{Other Scaling Relations}

\begin{figure}
\resizebox{\hsize}{!}{\includegraphics{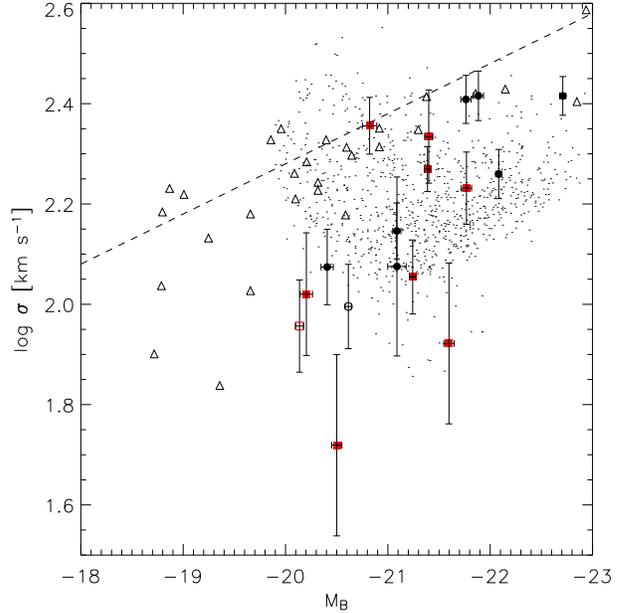}}
\caption{The Faber-Jackson relation in the rest--frame $B$--band
for our galaxies at $z$$\sim$1 and for
galaxies in Coma. Symbols are as in Fig.~\ref{FignMBvir}, small dots as in
Fig.~\ref{FigFP}, the dashed line is
the relation obtained at $z=0$ by Forbes \& Ponman (\cite{for99}).
}
\label{FigFJ}
\end{figure}

Figure~\ref{FigFJ} shows the relation between the velocity dispersion and the
absolute magnitude in the $B$ band -- the so--called Faber-Jackson
relation (Faber \& Jackson \cite{fab76}). 
Here the trends are less clear than with the FP, also because
even the low redshift objects show a large scatter.
However $z$$\sim$1 galaxies appear to be brighter than their
$z=0$ counterparts with the same velocity dispersion.

\begin{figure}
\resizebox{\hsize}{!}{\includegraphics{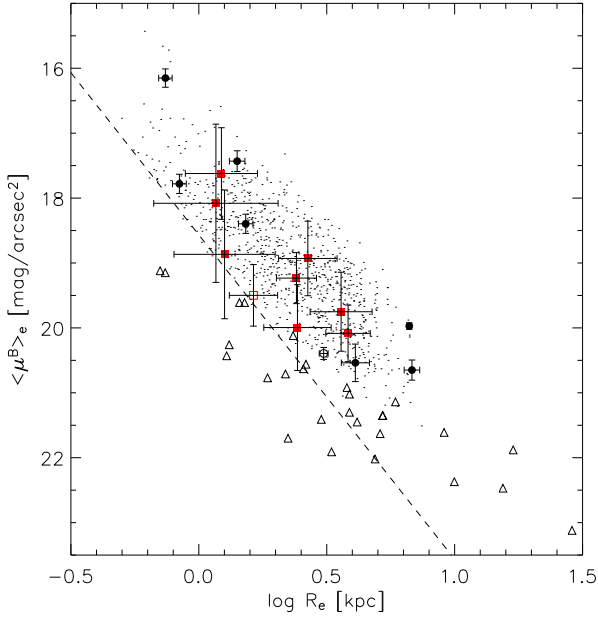}}
\caption{The Kormendy relation in the rest--frame $B$--band
for our galaxies at $z$$\sim$1 and for
galaxies in Coma. Symbols are as in Fig.~\ref{FignMBvir} and small dots as in
Fig.~\ref{FigFP}. The dashed line corresponds to $M_B=-20$.
}
\label{FigK}
\end{figure}

A clearer trend is shown by the purely morphological relation between 
the effective radius and the surface brightness, i.e. the Kormendy
(\cite{kor77}) relation (Fig.~\ref{FigK}). 
For a comparable size, $z$$\sim$1 early--type galaxies are about 1-2
magnitudes brighter in surface brightness than the low redshift ones.
This could in part be due to a selection effect: in fact Figure~\ref{FigK}
shows the limit for $M_B<-20$, which roughly corresponds to our apparent
magnitude limit of $R\la 24.4$ for measuring the velocity dispersion.
We also remark that the distribution of $R_e$
at $z$$\sim$1, although it partially overlaps the one for Coma, it is more
concentrated to smaller sizes.

\begin{figure}
\resizebox{\hsize}{!}{\includegraphics{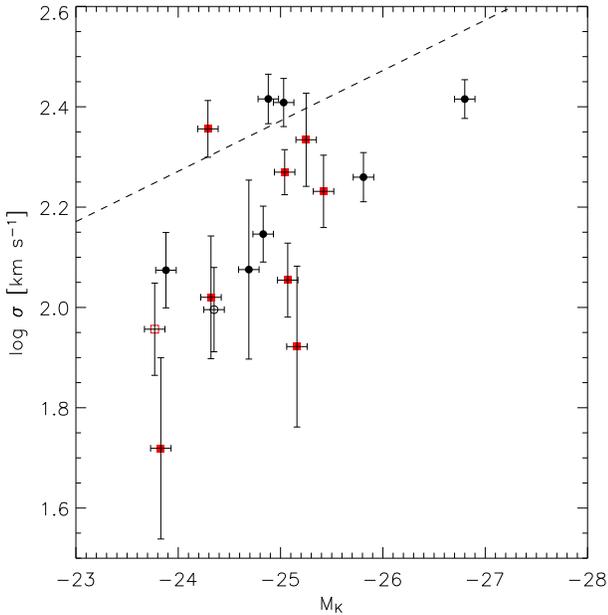}}
\caption{The Faber-Jackson relation in the rest--frame
$K$--band for our galaxies at 
$z$$\sim$1. Symbols are as in Fig.~\ref{FignMBvir}, the dashed line is
the local relation obtained for Coma cluster galaxies by Pahre,
Djorgovski and de Carvalho (\cite{pah98}).
}
\label{FigFJK}
\end{figure}

Using the $K$--band absolute magnitudes computed by Fontana et al.
(\cite{fon04}) by modest extrapolations of Bruzual \& Charlot
(\cite{bru03}) models fitted to the spectral energy distribution of our
galaxies, we have also obtained the Faber--Jackson relation in the
$K$--band (Fig.~\ref{FigFJK}). Most of our galaxies are brighter also in
the $K$--band than their local counterparts with the same velocity dispersion, 
indicating some evolution of the ${\cal M}/L_K$ ratio.

\subsection{Masses}

We return to the determination of dynamical masses by relaxing the
assumption of homology, which was used in equation \ref{mass}.
In fact we know that our galaxies are not homologous, since they have
different S\'ersic indices $n$, but seem to comply with a ``weak
homology'', as $n$ correlates with the luminosity (see
Fig.~\ref{FignMBvir}).
Bertin, Ciotti \& Del Principe (\cite{ber02}, BCD02 hereafter) have
expressed the dynamical mass of a spheroidal galaxy in the form:
\begin{equation}
{\cal M}_{{\rm K}_{\rm V}} = {{\rm K}_{\rm V}}\frac{\sigma^2 R_e}{G},
\label{massKv}
\end{equation}
where \kv~ is a virial coefficient which takes into account the specific
density distribution
of both luminous and dark matter (DM), the specific star orbit distribution,
and
projection effects (see also Lanzoni \&
Ciotti \cite{lan03}). By assuming spherical symmetry, global isotropy of the
velocity
dispersion tensor, and absence of DM, or a DM distribution
which
exactly parallels the stellar one, for a S\'ersic profile, \kv~ only
depends
on the S\'ersic index $n$ and on the aperture within which the velocity
dispersion $\sigma$ has been measured. For instance, the commonly used
factor 5 of equation \ref{mass}
is appropriate for a de Vaucouleurs $R^{1/4}$ profile and for aperture
radii of the order of $R_e/10$. An accurate fit of \kv~
for $1\le n\le10$
and small apertures ($R_e/8$) can be found in BCD02.
We have recomputed it for the 1.7 kpc aperture to which we correct our 
velocity dispersion measurements. The resulting values of
\kv~ and of ${\cal M}_{{\rm K}_{\rm V}}$ are listed in Table~\ref{sample}.
We remark that
this dynamical mass is larger by a factor of up to 1.55 than that
estimated using equation \ref{mass}.
We emphasize the existence of an elliptical galaxy, i.e. CDFS\_00633, at 
z=1.0963 with red colors, a $R^{1/4}$ profile, no line emission, probably
very low extinction, and a dynamical mass of $3.5\times 10^{11} {\cal M}_{\sun}$.

\begin{figure}
\resizebox{\hsize}{!}{\includegraphics{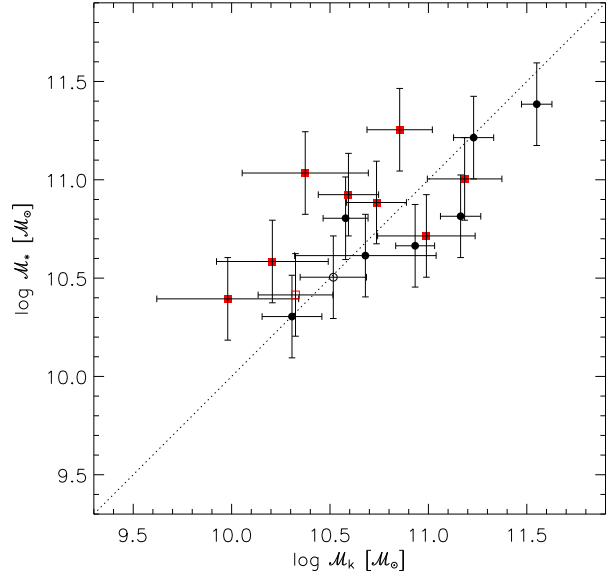}}
\caption{The comparison of dynamical and stellar masses for our sample at
$z$$\sim$1. Symbols are as in Fig.~\ref{FignMBvir}.
}
\label{FigMdynMst}
\end{figure}

We have compared our dynamical masses obtained with equation \ref{massKv}
with the stellar masses obtained by 
Fontana et al. (\cite{fon04}) for the K20 galaxies using their Best Fit model, 
where the spectrum best fitting the complete $UBVRIzJK_s$ multicolor
photometry for Salpeter IMF is used. However since the Salpeter IMF is known to
be inadequate (see e.g. Renzini \cite{ren05} and Bruzual \& Charlot 
\cite{bru03}) and all empirical determinations of the IMF indicate that
its slope flattens below $\sim 0.5\; {\cal M}_{\sun}$
(Kroupa \cite{kro01}, Gould et al. \cite{gou96}, Zoccali et al.
\cite{zoc00}), we have divided
the stellar masses of Fontana et al. (\cite{fon04}) by a factor of 1.72,
which is the age--averaged correction for a Chabrier (\cite{cha03}) IMF.
The comparison is shown in Fig.~\ref{FigMdynMst}.

The agreement between dynamical and stellar masses is good for the high
mass galaxies, but most of our lower mass galaxies have stellar masses
larger than the dynamical ones, with the maximum difference reaching a
factor of four.
Similar results have been recently found for lower redshift galaxies
by Drory, Bender \& Hopp (\cite{dro04}),
who however attribute the discrepancy to the limited reliability of the SDSS
velocity dispersion measurements for low masses galaxies.  We also remark
that, given our large spectroscopic aperture (in units of $R_e$), in case of
substantial rotational support, the adopted normalization of the observed
velocity dispersion (Sect. 3.2) might be incorrect, and the dynamical masses
of the stellar component underestimated (Riciputi et al. \cite{ric05}). This might
be the case particularly for the lower mass objects, that are also those showing
a faster evolution in the ${\cal M}/L_B$ ratio (see Section 5.1 and 
Fig.~\ref{FigMLMdyn}). 

\subsection{Comparison with previous results}

Van der Wel et al. (\cite{van05}), who have also analysed the FP for
galaxies in the CDFS, have two objects in common with us: their objects
CDFS--18 and CDFS--19 are our CDFS\_00633 and CDFS\_00571 respectively.
Our results are in reasonable agreement with theirs. More specifically,
considering ours vs. their results for CDFS\_00633 and for CDFS\_00571,
the $R_e$ is $6.65\pm0.12$ vs. $4.17\pm0.3$ kpc and $6.81\pm0.48$ vs.
$2.69\pm0.2$ kpc, the velocity dispersion is $260\pm23$ vs. $324\pm32$
km s$^{-1}$ and $182\pm21$ vs. $229\pm35$ km s$^{-1}$, the $K$ magnitude
is 16.94 vs. 16.98 and 17.54 vs. 17.53, the $U-B$ color is 0.54 vs. 0.32
and 0.26 vs. $-0.32$, the derived dynamical mass in
$10^{11} {\cal M}_{\sun}$ is $3.5\pm0.7$ vs. $5.1\pm1.0$ and $1.7\pm0.4$
vs. $1.7\pm0.5$ (we have used here the masses derived with equation
\ref{mass}, which is also assumed by van der Wel et al. \cite{van05}). 

This comparison suggests that our effective radii are somewhat larger than theirs,
particularly for CDFS\_00571.
Our velocity dispersions 
and dynamical masses are comparable with theirs. Surprisingly they
have much bluer colors than us, particularly for CDFS\_00571, for which
their value seems too blue, and in fact it is not plotted in their Fig. 8.

With respect to the morphological parameters obtained for the K20 galaxies in
the CDFS field by Cassata et al. (\cite{cas05}), we have refined those
previous values on a much smaller sample of objects by interactive
optimization of fits of individual galaxies, masking
out nearby objects and adjusting the local sky level.
In general Cassata et al. (\cite{cas05}) also find a decrease of the galaxy
size with redshift, particularly for the elliptical galaxies.

Also Daddi et al. (\cite{dad05}) have obtained morphological parameters
for 7 luminous early--type galaxies with $1.4<z<2.5$ using S\'ersic
profile fits with GALFIT on very deep HST+ACS images. They obtain
small effective radii, even slightly smaller than ours, and a wide range of
S\'ersic indices: $1<n<10$.

Finally, in comparing our relatively small sizes to the sizes obtained by 
others it is important to note that our spectroscopic galaxy type selection 
favours compact objects, while a morphological type selection,
such as e.g. that of Treu et al. (\cite{tre05a} and \cite{tre05b}), favours 
brighter and therefore probably larger objects.

\section{Discussion}

\subsection{Spheroids Evolution}

The observed offset of the $z$$\sim$1 FP from the local one is the result of the
evolution of one or more physical properties of spheroids. This could be either the
${\cal M}/L$ ratio, as in the classical analysis, or the size $R_e$, or a
combination of these two. Some evolution in the luminosity is actually 
expected as constituent stars grow older, even in the most passive
scenario (see the models in Fig.~\ref{FigDeltaML}). However there is also
evidence  of size evolution both from our data and from others (see 
Section 4.1). It is however not clear what is the exact relative weight of
these two contributions to the evolution of spheroids, and
how does an individual object evolve in size.

\begin{figure}
\resizebox{\hsize}{!}{\includegraphics{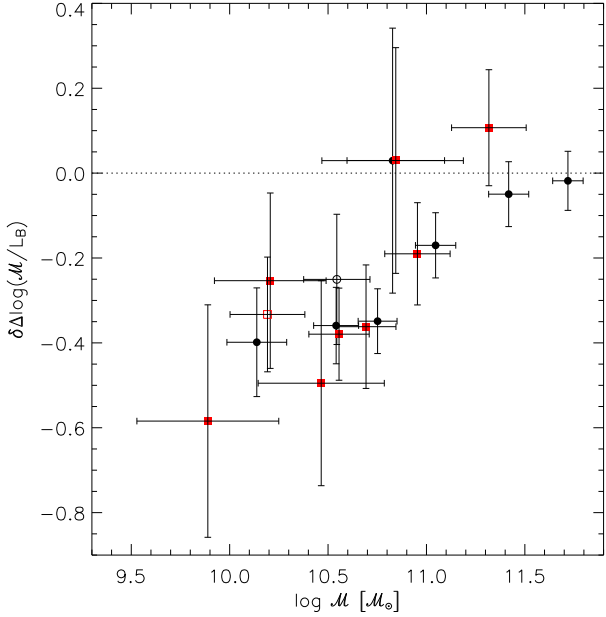}}
\caption{The differential evolution of the ${\cal M}/L_B$ ratio 
for each individual galaxy of our sample with respect to that of
old cluster massive
galaxies at the same redshift (see text). Symbols are as in
Fig.~\ref{FignMBvir}.
}
\label{FigMLMdyn}
\end{figure}

The observed rotation of the FP implies that the evolution of spheroids 
is not the same for all of them, whatever is the evolving parameter. 
Some have argued that the evolution could be mass dependent in
the sense that lower mass galaxies evolve later than the massive
ones (Fontana et al. \cite{fon04}; Pozzetti et al. \cite{poz03}). Very
recently Treu et al. (\cite{tre05a}) and van der Wel et al. (\cite{van05}) have
analysed the results of kinematic studies of field early--type galaxies
with $0.2<z<1.2$ in term of a mass dependent evolution. 
In Section 4.2 we have already shown that the ${\cal M}/L_B$ evolution of
our early--type galaxies depends on their dynamical mass. This is clearer
in Figure ~\ref{FigMLMdyn}, where the difference between the ${\cal
M}/L_B$ evolution of our galaxies and the fit derived by van Dokkum \&
Stanford (\cite{van03}) for the cluster massive galaxies ($\delta\Delta
{\rm log}({\cal M}/L_B) = \Delta {\rm log}({\cal M}/L_B) + 0.46z$) is
plotted as a function of mass. The dependence of the differential ${\cal M}/L_B$
evolution on mass appears to have a threshold around $10^{11}{\cal
M}_{\sun}$, in the sense that more massive galaxies evolve in a 
way similar to that of cluster massive galaxies, while less massive objects 
evolve progressively faster and their ${\cal M}/L_B$ ratio is lower than
that of cluster massive galaxies by a factor of up to about 3 for galaxy
masses of the order of $10^{10}{\cal M}_{\sun}$. We emphasize that the
differential ${\cal M}/L_B$ evolution, clearly visible in Figure
~\ref{FigMLMdyn} and in similar figures by Treu et al. (\cite{tre05a}) and
by van der Wel et al. (\cite{van05}), implies that the FP must rotate with 
redshift (Renzini \& Ciotti \cite{ren93}).

The differential ${\cal M}/L_B$ evolution which we observe is similar to the
so--called ``downsizing'' (Cowie et al. \cite{cow96}, see also Kodama et al.
\cite{kod04}, Treu et al. \cite{tre05a}), i.e. the later and/or longer
lasting formation of lower mass galaxies.
The reason for such later evolution of less massive spheroids could be
found in
a change with mass of either their stellar populations (IMF, age, metallicity),
or of some structural/dynamical parameter, such as the
DM distribution, the density profile, the degree of
anisotropy and the partial rotational support, or a combination of these.
However the only structural/dynamical changes possible during the Universe
lifetime are those induced by merging.

\begin{figure}
\resizebox{\hsize}{!}{\includegraphics{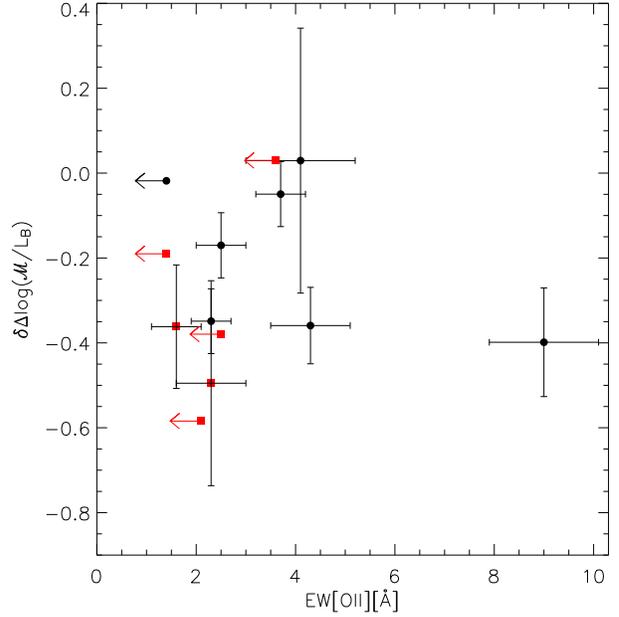}}
\caption{The dependence of the differential evolution in ${\cal M}/L_B$ on the 
rest--frame equivalent width of the [OII]3727 emission line for our
spheroidal galaxies at $z$$\sim$1. Symbols are as in Fig.~\ref{FignMBvir}.
}
\label{FigEWOII}
\end{figure}

\begin{figure}
\resizebox{\hsize}{!}{\includegraphics{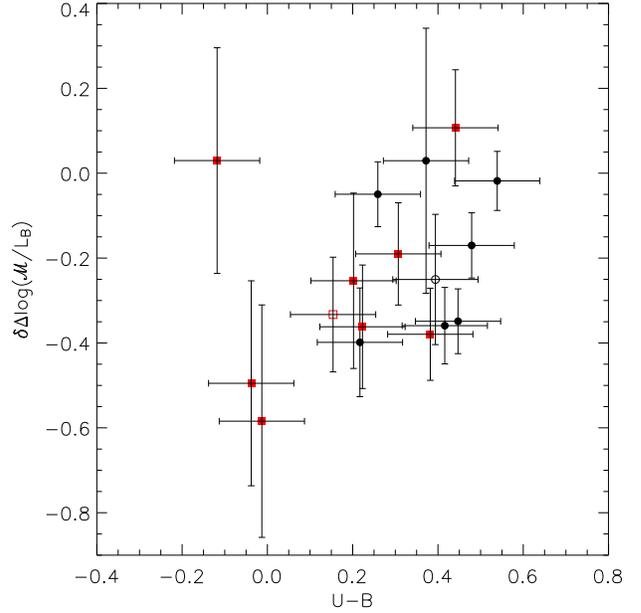}}
\caption{The dependence of the differential evolution in ${\cal M}/L_B$
on the rest--frame $U-B$ color. Symbols are as in Fig.~\ref{FignMBvir}.
}
\label{FigU_B}
\end{figure}

A testable hypothesis is that the lower mass--to--light ratio of the less
massive galaxies is due to recent star formation or younger ages 
increasing their $B$--band
luminosity. We have therefore looked at the dependence
of $\delta\Delta {\rm log}({\cal M}/L_B)$ on star formation indicators as
the [OII] equivalent width and the rest--frame $U-B$ color. 
The dependence on the [OII] equivalent width (Fig.~\ref{FigEWOII})
does not show any correlation, indicating that the lower ${\cal M}/L_B$
ratio is not due to on--going star formation activity. This is actually
consistent with the fact that our galaxies have been selected as having
early--type spectra and that the detected [OII] line fluxes correspond to a
rather low star formation rate, ranging
between 0.10 and 0.18 ${\cal M}_{\sun}$ yr$^{-1}$, using the conversion of
Kewley, Geller \& Jansen (\cite{kew04}) and assuming no extinction.
On the other hand a correlation exists between the differential evolution in 
${\cal M}/L_B$ ratio and the $U-B$ color (Fig.~\ref{FigU_B}), with galaxies
with lower ${\cal M}/L_B$ ratio being bluer, indicating that indeed star
formation activity has been proceeding until a rather recent past in the
lower mass galaxies.

As anticipated in Sect. 4.4, the combined effect of substantial
rotational support and large spectroscopic apertures in the 
low-mass galaxies (e.g., Davies et al. \cite{dav83}), could produce a trend
similar to the differential evolution. In fact, aperture velocity
dispersion of isotropic rotators, measured at or beyond $R_e$, can be
significantly lower than the same quantity measured in the central
region. 
Thus, the empirical correction of J\o rgensen et al.
(\cite{jor95b}; see Sect. 3.2),
that has been locally determined from velocity dispersion
profiles obtained through small spectroscopic apertures (where the effect
of rotation is negligible), might not be appropriate in our case (large
apertures) if rotation is important.
Unfortunately, a
quantitative estimate of such effect is not easy; however, a detailed
analysis of rotational and flattening effects on aperture velocity
dispersion of axysimmetric galaxy models shows that the variation is of 
the order of $\approx$ 30\% for apertures of $\approx R_e$ (Riciputi et
al. \cite{ric05}; it increases up to $\approx$ 80\% when the whole galaxy image
is taken into account, as assumed by Bender, Burstein \& Faber \cite{ben92}),
much less than our estimated differential evolution. Note however that
the correction could be larger if a rotating disk is also present (see
CDFS\_00571 in Fig.~\ref{Figmosa_sp1}). In order to check the importance of this effect on 
our data, we have looked for signs of rotation on the two-dimensional
spectra of our galaxies
with line emission. Unfortunately the emission lines are too faint and
compact to possibly show any sign of rotation at the S/N ratio of our
spectra.

Finally, we remark that data in Fig.~\ref{FigMLMdyn} were obtained under the assumption 
of strong $R^{1/4}$ homology (Eq. \ref{mass}), and so it is natural to ask how
the same figure is modified by adopting instead Eq. \ref{massKv}. In fact we
know that our fainter galaxies have smaller S\'ersic indices (see
Fig. ~\ref{FignMBvir}), and, since low $n$ correspond to high values of \kv, this may reduce
the predicted differential evolution. However the {\it maximum} effect due to
non-homology is of the order of 55\% (see Sect. 4.4), and therefore,
as in case of rotation, much smaller than the required factor of 3. 

Of course, the two effects above could combine and reduce the amount
of the required intrinsic differential evolution, but it seems to
be impossible to blame just systematic effects for the behavior of
data in Fig.~\ref{FigMLMdyn}, for which younger mean ages for low mass
galaxies probably have a major effect (Fig.~\ref{FigU_B}).

\subsection{Comparison with Hierarchical Models Predictions}

One of the currently open questions of observational cosmology is whether
the
observed evolution of early--type galaxies is consistent with the
predictions of the hierarchical theories of galaxy formation. Our sample of
early--type galaxies, being selected in the $K$ band, which is particularly
sensitive to galaxy mass, and being at the highest redshifts for which
kinematic information has been obtained so far, is particularly well
suited for this check.

If the early--type galaxies form by merging, as predicted by the
hierarchical models, then we should see a different evolution in the
cluster and in the field environment, where the merging probabilities can
be different. We fail to observe such effect, since our most massive
field spheroids evolve at a rate indistinguishable from that of massive
spheroids in clusters, and in the lower mass range there appear to exist
early-type galaxies in clusters which evolve at a faster rate,
comparable to that of our less massive
field galaxies. Examples of this are our two galaxies at
$z=0.67$, which definitely evolve faster than cluster massive galaxies,
although they are also likely in clusters. We note that also the two galaxies 
with ${\cal M} < 10^{11}{\cal M}_{\sun}$ in fig. 6a of van Dokkum \& Stanford
(\cite{van03}) and the two lowest mass E+A galaxies observed by van
Dokkum et al. (\cite{van98}) in a cluster at $z=0.83$ appear to have a lower ${\cal
M}/L$ ratio than the massive galaxies in the same cluster.
We do not exclude that the evolution of spheroids may also depend on the
environment, but to test this hypothesis it is necessary to study the FP on
well selected and complete samples of {\it cluster} early-type galaxies,
overcoming the obvious observational effect, that usually prevents less
luminous galaxies from finding empty slits in multi-object spectroscopic
observations of high density cluster fields.

If, as predicted by some versions of the hierarchical merging model, the
stellar populations of more massive galaxies are appreciably younger than
those of smaller galaxies, the FP should become flatter with increasing
redshift (e.g., Renzini \cite{ren99}).
This is contrary to the observed steepening of the FP, and suggests that, if
spheroids form by merging, the building blocks of the most massive objects
at $z\sim 1$ should have formed and assembled well before those constituting
the lower mass galaxies at the same redshift.

To further check the ability of the hierarchical scenario in describing how
galaxies form and evolve in the Universe, we have compared our results with
the predictions of GalICS (Hatton et al. \cite{hat03}), a hybrid (N-body and
semi-analytic) hierarchical model of galaxy formation, that matches rather
well the redshift distribution of the K20 survey (Blaizot et al.
\cite{bla05}).
To perform a proper comparison with the observations, we have selected all
field elliptical and S0 galaxies with $K_s< 20$ and $R<24.4$ in 3 timesteps,
corresponding to redshifts $z=0.87, 1.01$ and 1.23.
Although semi--analitic models are unable, by construction, to resolve the
internal structure and dynamics of galaxies, spheroidal components in GalICs
are approximated as Hernquist (\cite{her90}) spheres, characterized by a
half--mass
radius $r_h$ and a velocity dispersion at $r_h$ (see Hatton et al.
(\cite{hat03})
for details). We have therefore used the Hernquist model to correct these
values
and derive the corresponding effective radius $\re$ and the one-dimensional
velocity dispersion within a $1.19\,h^{-1}$ kpc aperture, consistently with
what done for the observations. The resulting values and the corresponding
effective surface brightness have been used to build the simulated scaling
relations, that are shown in Figures~\ref{FigFP}, \ref{FigFJ} and
\ref{FigK}.
The predicted FP appears
to be similar to the observed one, even if its scatter is larger than
observed. A good agreement is found for the Kormendy relation, while the
discrepancies at the faint end of the Faber--Jackson relation suggest
excessively large velocity dispersions and/or underestimated
$B$--luminosities
for simulated objects with $M_B\la -21$. The model stellar masses, instead,
are found to be in the correct range, between $10^{10}$ and
$2\times10^{11}\msol$.  If one considers that the definition of
morphological
types in the model (based on the bulge-to-disc luminosity ratio in the B
band) is different from the one adopted in the observations, and that
structural and dynamical properties of galaxies can be only roughly
described
in this kind of modelling, the present comparison can be considered quite
satisfactory.  However, a major problem exists concerning the colours of
simulated galaxies, which are too blue compared to the observations: for the
selected GalICS sample, the rest-frame $B-K$ ranges between 1.9 and 4.6,
with
median at $\sim 3$, while all observed galaxies have $B-K>3.2$. In
addition, excessively large dust extinctions are needed to obtain the
reddest
colours in the model galaxies.
This is a well known, but still unsolved, problem common to all
semi--analytic
and hybrid models of this kind (see, e.g., Firth et al. \cite{fir02}; Benson
et
al. \cite{ben02}; Pozzetti et al. \cite{poz03}; Somerville et al.
\cite{som04}),
suggesting that a
re--examination of how star formation proceeds in galaxies is required,
particularly at high redshift.

\section{Conclusions}

We have obtained photometric (total magnitudes in various bands), morphological
(effective radius, average surface brightness within the effective radius,
S\'ersic index) and kinematic (velocity dispersion) parameters for an
almost complete sample of 15 field early--type galaxies selected from the K20
survey in the redshift range $0.88\leq z \leq 1.30$. Apart from
completeness, our sample has the advantage of being selected in the
$K$--band, of having the galaxy type assigned spectroscopically, and of
covering two separate fields. 

From these data we obtain the following results:

1. The distribution of effective radii is shifted towards smaller sizes
compared to the local one.

2. S\'ersic indices span a range between 0.5 and 4, and correlate with
$B$--band luminosity.

3. The resulting FP in the rest--frame $B$--band at $z$$\sim$1 is offset from,
and steeper than, the local one, but keeps a remarkably small scatter.

4. Under the assumption of homology, and at fixed $R_e$, the evolution of 
the dynamical ${\cal M}/L_B$ ratio shows a dependence on galaxy mass: for our
more massive objects, it is similar to that of {\it cluster  massive
galaxies}, while it is faster for the less massive ones, with a threshold
around $10^{11}{\cal M}_{\sun}$.

5. For a given velocity dispersion or size some of the $z$$\sim$1 galaxies are
considerably brighter than local ones, both in the rest--frame $B$-- and
$K$--band.

6. Dynamical masses (properly obtained by taking into account the observed
S\'ersic index) are consistent with stellar
masses (derived from model fits to the spectral energy distribution) in
the high mass range, while they are smaller in the low mass range.

7. In about half of our $z$$\sim$1 early-type galaxies we have detected [OII]
line emission, corresponding to a low star formation rate ranging from 0.10
to 0.18 ${\cal M}_{\sun}$ yr$^{-1}$.

We give the following interpretation:

1. Spheroidals evolve in ${\cal M}/L$ ratio, and possibly also in size.

2. The rate of evolution over the last 2/3 of the Universe
lifetime depends on the galaxy mass, low mass galaxies evolving faster.

3. A fraction of the detected differential evolution might be due to a
stronger rotational support in low mass galaxies with respect to the high mass 
ones, and to the fact that observational data are routinely interpreted under
the assumption of strong homology. However the correlation with galaxy color
indicates that a significant fraction of the detected differential
evolution is indeed intrinsic, i.e. due to more recent star formation
episodes (the so called ``downsizing'').

4. We fail to detect differences in the evolution with the environment,
and the rotation of the FP goes in the opposite sense than predicted by the
hierarchical merging models.

5. The GalICS hierarchical galaxy formation model 
is able to produce galaxies which are
in the correct mass and size range and reproduce the various scaling 
relations, although with a larger scatter for the FP and with
unrealistic colours and dust extinctions.

Since neither homology (constant \kv) nor the other extreme case of weak 
homology (constant mass--to--light ratio, BCD02) appear to be holding for our 
$z$$\sim$1 spheroids, we are astonished that the FP keeps such a small scatter,
while shifting and rotating as the Universe evolves,
thus posing additional and more stringent challenge to
our understanding of galaxy formation and evolution.

\begin{acknowledgements}

We are grateful to Julien Devriendt, George Djorgovski,
Gianni Fasano, Alberto Franceschini and Tommaso Treu for helpful
discussions. We thank Rino Bandiera, Jaron Kurk and Marco Scodeggio
for useful software advice,
Adriano Fontana for providing the stellar masses, and the referee for
constructive comments.
This research has made use of NASA's Astrophysics Data
System and of the SIMBAD database, operated at CDS, Strasbourg, France.

\end{acknowledgements}

\end{document}